\documentclass[10pt,twocolumn,letterpaper]{article}

\usepackage{cvpr}
\usepackage{times}
\usepackage{epsfig}
\usepackage{graphicx}
\usepackage{amsmath}
\usepackage{amssymb}
\usepackage{longtable}
\usepackage{paralist}
\usepackage{multirow,epstopdf,subfigure}
\usepackage{enumitem}

\usepackage{booktabs}
\usepackage{algorithm}
\usepackage{algorithmic}

\usepackage{newfloat}
\usepackage{listings}

\usepackage{epsfig}
\usepackage{amsmath}
\usepackage{amssymb}
\usepackage{longtable}
\usepackage{paralist}
\usepackage{multirow,epstopdf,subfigure}
\usepackage{enumitem}
\usepackage{color}

\usepackage[breaklinks=true,bookmarks=false]{hyperref}

\cvprfinalcopy 

\def\cvprPaperID{****} 

\begin{document}

\title{Self-Verification in Image Denoising}

\author{Huangxing Lin\textsuperscript{1}, \  Yihong Zhuang\textsuperscript{1}, \ Delu Zeng\textsuperscript{2}, \  Yue Huang\textsuperscript{1}, \ Xinghao Ding\textsuperscript{1*}, \  John Paisley\textsuperscript{3}\\
\normalsize \textsuperscript{1}Xiamen University, \normalsize \textsuperscript{2}South China University of Technology, China\\
\normalsize \textsuperscript{3} Columbia University,  USA\\
{\small \textsuperscript{*}\tt Corresponding author: dxh@xmu.edu.cn}
}

\maketitle


\begin{abstract}

	We devise a new regularization, called self-verification, for image denoising. This regularization is formulated using a deep image prior learned by the network, rather than a traditional predefined prior. Specifically, we treat the output of the network as a ``prior'' that we denoise again after ``re-noising''. The comparison between the again denoised image and its prior can be interpreted as a self-verification of the network's denoising ability. We demonstrate that self-verification encourages the network to capture low-level image statistics needed to restore the image. Based on this self-verification regularization, we further show that the network can learn to denoise even if it has not seen any clean images. This learning strategy is self-supervised, and we refer to it as Self-Verification Image Denoising (SVID). SVID can be seen as a mixture of learning-based methods and traditional model-based denoising methods, in which regularization is adaptively formulated using the output of the network. We show the application of SVID to various denoising tasks using only observed corrupted data. It can achieve the denoising performance close to supervised CNNs.

\end{abstract}

\section{Introduction}

\noindent Images captured by various devices are prone to noise and corruption due to limited imaging environments such as low light and slow shutter speed. Denoising is the problem whereby we recover the underlying clean image from its noise measurements. While sometimes the purpose may be for aesthetic reasons, downstream computer vision tasks, such as detection and segmentation, are also greatly facilitated by having minimally corrupted inputs. 

A noisy image $y$ is usually modeled as
\begin{equation}
\label{eq.y}
y=x+n,
\end{equation}
where $n$ represents the noise, $x$ is the clean image to be restored. This inverse problem is challenging because the statistics of $n$ are usually unknown and often complex.

In general, model-based image denoising algorithms aim to solve this problem through optimizing functions of the form
\begin{equation}
\label{eq.map}
x^* = \arg \mathop {\min }\limits_x \left\| {x - y} \right\|_2^2 + \alpha R(x),
\end{equation}
where $\left\| {x - y} \right\|_2^2$ is a data fidelity term that ensures the solution agrees with the observation, $R(x)$ is a regularizer and $\alpha$ is a trade-off parameter. The first challenge of Eq. (\ref{eq.map}) is choosing $R(x)$, which encodes the pre-known image properties and directs the solution towards a more plausible image. (Given this sense of $R(x)$, we refer to it as a ``prior'' although we are not taking a Bayesian perspective.) Some common priors for constructing $R(x)$ include total variation, non-local self-similarity, and others \cite{dong2015low, meng2013robust}. However, these preselected priors often do not model the finer properties of natural images, and so often lead to degraded results.

\begin{figure}[t]
	\centering

	{\includegraphics[width=3in]{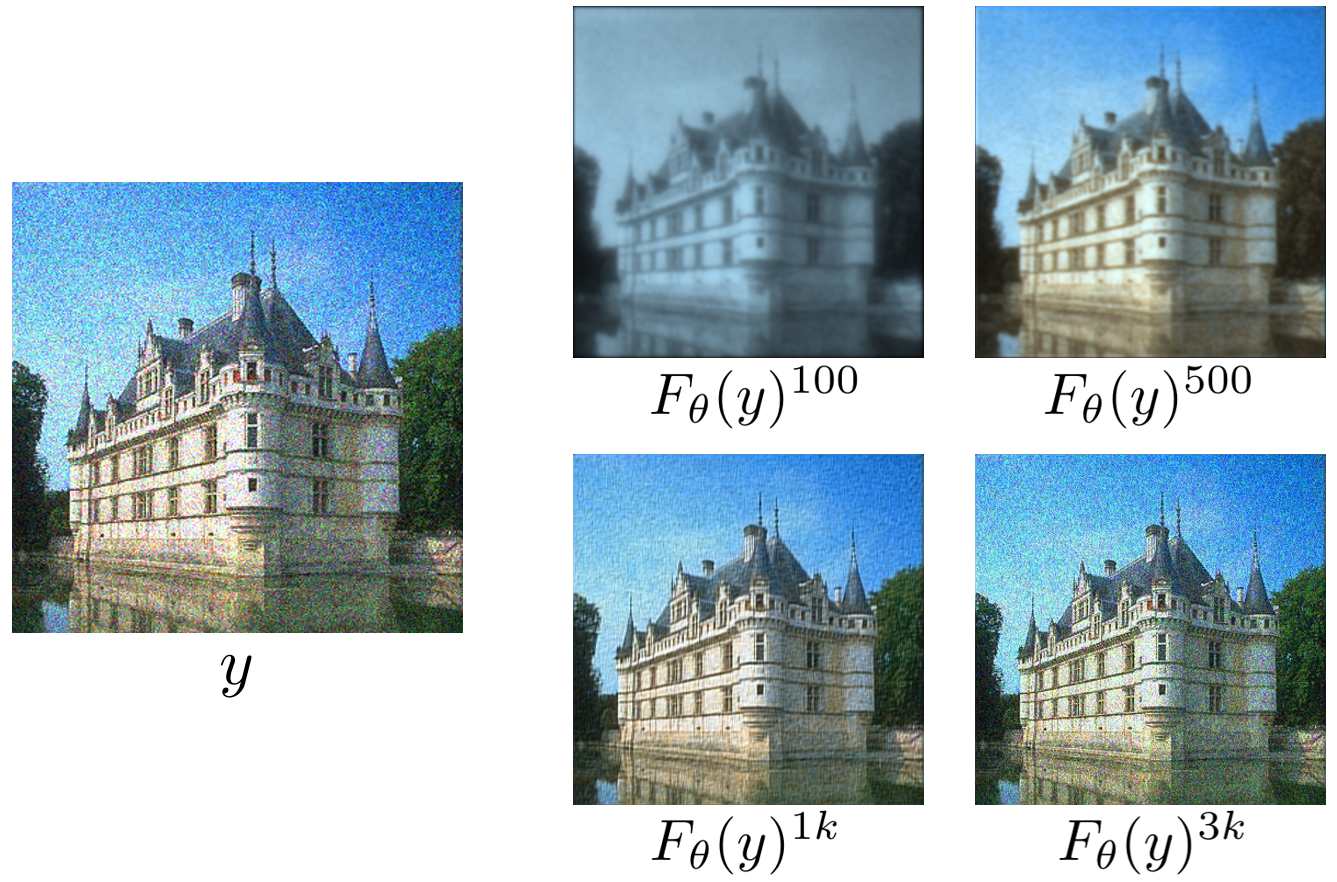}}
	
	{\tiny }
	
	\caption{Visual results obtained by minimizing Eq. (\ref{eq.dip2}). The superscript represents the number of iterations in training.}
	
	\label{fig0}
\end{figure}

Recently, supervised deep networks have achieved unprecedented success in image denoising by learning image priors and noise statistics from pairs of noisy/clean images \cite{zhang2019residual}. Most CNN denoisers, \textit{e.g.} DnCNN \cite{zhang2017beyond} and VDN \cite{yue2019variational}, have superior performance over model-based denoisers. Unfortunately, in most cases, collecting a large number of realistic paired data is difficult, which limits the application of supervised CNNs. This leads to the topic of learning to denoise without paired data, the common approach of non-deep models, but less addressed by neural networks. Some self-supervised methods \cite{laine2019high} show that only noisy data is required to train a denoising network under certain minor assumptions (\textit{e.g.} noise is zero-mean). One of these methods is called a deep image prior (DIP) \cite{ulyanov2018deep}, which has attracted much attention because it handles image denoising from a unique perspective. DIP found that the structure of CNN is naturally an implicit regularizer for image denoising. Specifically, DIP uses a deep CNN to reparameterize a noisy image $y$, \textit{i.e.},
\begin{equation}
\label{eq.dip}
\mathop {\min }\limits_\theta  \left\| {{F_\theta }(z) - y} \right\|_2^2,
\end{equation}
where $z$ is a fixed random vector, ${F_\theta }( \cdot )$ represents the deep CNN and $\theta$ is its parameters. The optimization trajectory of Eq. (\ref{eq.dip}) will pass through a good local optimum (\textit{i.e.} a clean image) before overfitting to $y$. So denoising can be completed by stopping training early. Moreover, we found that similar results (see Figure \ref{fig0}) can be observed even if $z$ in Eq. (\ref{eq.dip}) is replaced by $y$,
\begin{equation}
\label{eq.dip2}
\mathop {\min }\limits_\theta  \left\| {{F_\theta }(y) - y} \right\|_2^2.
\end{equation}

The effectiveness of DIP relies on the fact that the CNN naturally tends to capture low-level image statistics, while showing strong ``impedance'' to unstructured noise \cite{ulyanov2018deep}. DIP demonstrates the potential of neural networks to learn image priors without clean labels.

\begin{figure*}[t]
	\centering
	\includegraphics[width=5.2 in]{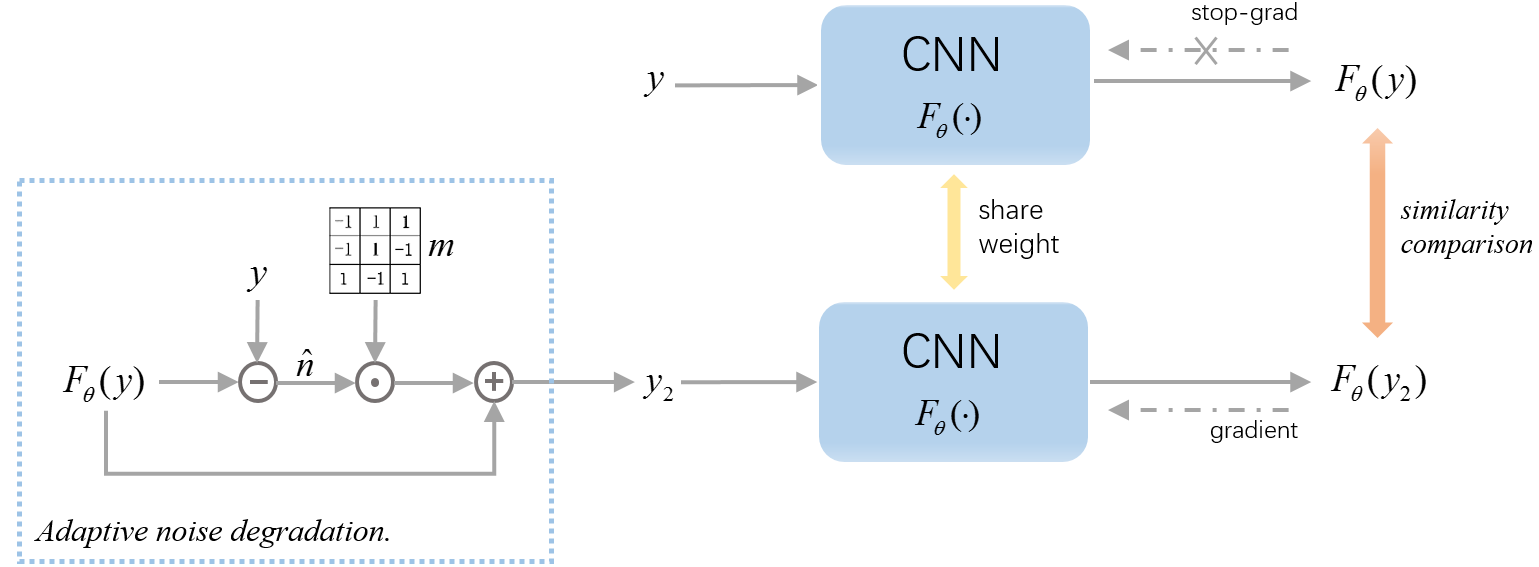}

	\caption{Illustration of self-verification image denoising (SVID). Self-verification consists of two steps: adaptive noise degradation and similarity comparison. Self-verification refers to using the output of the network to verify its own denoising ability. The dimensions of $m$ and $\hat{n}$ are the same. For better visualization, only nine elements in $m$ are shown.}
	
	\label{fig_svid}
\end{figure*}

\subsection{Motivation}

While DIP is an effective unsupervised approach, a drawback for real-world applications is its time-consuming training that requires human intervention for stopping. Inspired by DIP, in this paper we explore a more flexible use of the neural network image prior. Similar to DIP, we only focus on self-supervised cases; the difference is that our goal is to formulate an explicit regularization for denoising using the prior determined by the network.

We use a deep CNN to perform the denoising task in Eq. (\ref{eq.map}), so Eq. (\ref{eq.map}) can be rewritten as
\begin{equation}
\label{eq.map_cnn}
\theta^*  = \arg \mathop {\min }\limits_\theta  \left\| {{F_\theta }(y) - y} \right\|_2^2 + \alpha R({F_\theta }(y)).
\end{equation}

Since DIP has confirmed the affinity of the deep network for learning low-level image statistics, we expect that this network will first capture low-level image statistics as well in early iterations before attempting to memorize the noise when solving Eq. (\ref{eq.map_cnn}). Since the output of the network (\textit{i.e.} ${F_\theta }(y)$) contains the low-level statistics required to restore the image, we consider using ${F_\theta }(y)$ as the prior for image denoising. Put another way, can the output of the denoising network be used to construct a regularization for itself? In this paper, we give a feasible solution for performing this task.

\subsection{Our Proposed Contribution}

Suppose there is a second noise-degraded image $y_2$ in addition to $y$ for which both share the same latent clean image $x$, and which have independent and identically distributed (i.i.d.) noise. Based on $y_2$ and $y$, we devise a new regularization term for image denoising. Then, we could consider for Eq. (\ref{eq.map_cnn}) the function
\begin{equation}
\label{eq.new_reg}
\theta^*  = \arg \mathop {\min }\limits_\theta  \left\| {{F_\theta }(y) - y} \right\|_2^2 + \alpha \left\| {{F_\theta }({y_2}) - {F_\theta }(y)} \right\|_2^2.
\end{equation}

Since $y_2$ and $y$ have different noise, ${F_{\theta^*} }(y)=x$ is obviously a good candidate solution for minimizing Eq. (\ref{eq.new_reg}).
However, in most real-world applications, $y$ is available, but we do not have a second noisy image $y_2$. Instead, we use the output of the network to synthesize a new $y_2$,
\begin{equation}
\label{eq.A}
{y_2} \equiv D({F_\theta }(y)) = {F_\theta }(y)+n_2,
\end{equation}
where $D(\cdot)$ is a predefined adaptive noise degradation function and $n_2$ is newly generated noise. We discuss $D(\cdot)$ later. Combining Eqs. (\ref{eq.new_reg}) and (\ref{eq.A}), we have
\begin{equation}
\label{eq.self_reg}
\theta^* = \arg \mathop {\min }\limits_\theta  \left\| {{F_\theta }(y) - y} \right\|_2^2 + \alpha \left\| {{F_\theta }(D({F_\theta }(y))) - {F_\theta }(y)} \right\|_2^2.
\end{equation}

The first term of Eq. (\ref{eq.self_reg}) implicitly constrains the removed noise to be zero mean. In the second term of Eq. (\ref{eq.self_reg}), if the noise of $D({F_\theta }(y))$ and $y$ share similar statistical characteristics, but ${n_2} \ne n$, then ${F_{\theta^*} }(y)=x$ is still a good solution for Eq. (\ref{eq.self_reg}). We refer to the second term of Eq. (\ref{eq.self_reg}) as a self-verification regularization. In $\left\| {{F_\theta }(D({F_\theta }(y))) - {F_\theta }(y)} \right\|_2^2$, ${F_\theta }(y)$ can be regarded as a deep image prior learned by the network. The similarity comparison between ${F_\theta }(D({F_\theta }(y)))$ and ${F_\theta }(y)$ can be seen as using a prior generated by the network to verify the denoising behavior of the network itself. We update the network to minimize $\left\| {{F_\theta }(D({F_\theta }(y))) - {F_\theta }(y)} \right\|_2^2$. When $\theta$ converges and the network becomes a good denoiser, the discrepancy between ${F_\theta }(D({F_\theta }(y)))$ and ${F_\theta }(y)$ should be small.


With self-verification regularization, we further show how a deep CNN learns to denoise with self-supervision. We call this training method Self-Verification Image Denoising (SVID). SVID does not rely on other noise statistics except the assumption of zero-mean noise. To perform self-verification, SVID uses a Siamese network as the backbone. Although SVID is based only on noisy training data, its denoising results are impressive. We demonstrate the superior performance of SVID in various denoising experiments. SVID significantly outperforms other state-of-the-art self-supervised methods. 

%
%

\section{Related Work}
In this section, we review related topics most relevant to this work, including model-based denoising methods, learning-based denoising methods and Siamese networks.

Most model-based denoising methods focus on using handcrafted priors \cite{xu2018trilateral, buades2005non, meng2013robust}, \cite{zhao2014robust} to construct regularization. Over the past decades, various ideas have been proposed, all aiming to identify sources of inner structure in visual data. For example, the non-local self-similarity (NSS) prior \cite{yao2019nonconvex} is widely used in image denoising; some well-known NSS-based methods include BM3D \cite{dabov2007image} and WNNM \cite{gu2014weighted}. Other prominent techniques, such as wavelet coring \cite{simoncelli1996noise}, total variation \cite{selesnick2017total} and low-rank assumptions \cite{dong2015low} are also standard approaches.

In recent years, supervised deep learning with CNNs has shown excellent denoising performance \cite{zhang2021accurate}. Many methods adopt sophisticated network structures to achieve better denoising results  \cite{zhang2018ffdnet,yue2020dual,liu2018non}. Their effectiveness is due to their ability to learn image priors from large amounts of paired data \cite{zhang2020residual}. However, they require paired noisy/clean images. To mitigate this issue, Lehtinen \textit{et al.} \cite{lehtinen2018noise2noise} demonstrate that the denoising CNN can be trained with pairs of independent noisy measurements of the same scene. This training strategy is called Noise2Noise, and achieves denoising performance on par with general supervised learning methods.

To relax the supervised requirement, networks trained with only noisy images has recently been considered \cite{laine2019high}. For instance, Noise2Void \cite{krull2019noise2void} proposes a blind spot strategy, which trains a network to predict masked pixels using their neighbors so as to achieve denoising. The blind spot strategy has strong denoising ability, and it is adopted by many self-supervised methods, such as Self2Self \cite{quan2020self2self} and Noise2Self \cite{batson2019noise2self}. Later, Neighbor2Neighbor \cite{huang2021neighbor2neighbor} showed that two different subsamples of a noisy input can also be the training data for denoising.

As for the choice of network, Siamese networks are natural and effective tools for modeling invariance, and therefore have become a common structure for self-supervised representation learning methods \cite{chen2021exploring,zbontar2021barlow,grill2020bootstrap}. Generally, inputs are distorted versions of the same sample, and maximize the similarity subject to different conditions. In this paper, we also use a Siamese network, not to learn a robust representation, but to discover the clean target shared between the inputs $y$ and $y_2$.

\section{Methodology}
Given a noisy dataset $\Omega^{noisy}=\{y^i \}^N_{i=1}$, we aim to use these noisy images to learn to denoise without using other clean images. We propose a self-verification image denoising (SVID) approach, as shown in Figure \ref{fig_svid}. SVID is derived from Eq. (\ref{eq.self_reg}), in which self-verification regularization plays a major role.  Unlike DIP \cite{ulyanov2018deep} trains a new network for each image, we use all noisy images to train a shared denoising network by mini-batch gradient descent. For simplicity, we set the batch size to 1 in the objective function of SVID, but it is easy to expand to the case where the batch size is greater than 1. After training, the denoising CNN in SVID is directly applied to new noisy images without additional learning.


\subsection{Overview of SVID}
Self-verification consists of two steps: adaptive noise degradation and similarity comparison. We use a Siamese network, where the goal is first to input $y$ to a CNN to obtain a denoised image ${F_\theta }(y)$ and its removed noise $\hat n = y - {F_\theta }(y)$. Since low-level image statistics are easier to be captured by deep CNNs than noise, we assume that ${F_\theta }(y)$ is a clean image. We then synthesize $y_2$ by the following adaptive noise degradation function,
\begin{equation}
\begin{aligned}
\label{eq.y2}
{y_2} &\equiv D({F_\theta }(y))\\
&= {F_\theta }(y) + {n_2}\\
&= {F_\theta }(y) + m \odot \hat n\\
&= {F_\theta }(y) + m \odot (y - {F_\theta }(y)),
\end{aligned}
\end{equation}
where ${n_2} = m \odot \hat n$ is a new noise which is adaptively synthesized with respect to $\hat n$. $\odot$ represents element-wise multiplication, $m$ is a random binary ($1$ or $-1$) mask that is independent of $\hat n$ but has the same dimension as $\hat n$. In each iteration of learning, a new $m$ is sampled. Each element in $m$ is $1$ with a probability of $p=0.5$, and $-1$ otherwise.

Next, the network learns to denoise by maximizing the similarity between ${F_\theta }(y)$ and ${F_\theta }(y_2)$. This is achieved by solving the following optimization problem,
\begin{equation}
\begin{aligned}
\label{eq.loss}
\theta^*  = \arg \mathop {\min }\limits_\theta  \left\| {{F_\theta }({y_2}) - {\rm{Stopgrad}}({F_\theta }(y))} \right\|_2^2.
\end{aligned}
\end{equation}
``Stopgrad'' means that the gradient does not backpropagate to ${F_\theta }(y)$, which helps prevent trivial solutions (\textit{i.e.} ${F_\theta }( \cdot ) = c$, $c$ is a constant.). The stop-gradient procedure is a common strategy for learning Siamese networks \cite{chen2021exploring}.

\begin{figure}[t]
	\centering
	
	{\includegraphics[width=0.6in]{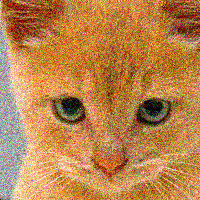}}
	{\includegraphics[width=0.6in]{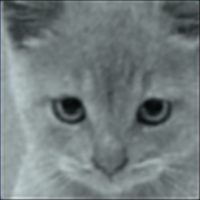}}
	{\includegraphics[width=0.6in]{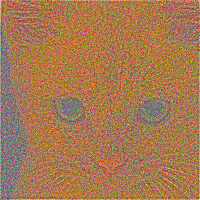}}
	{\includegraphics[width=0.6in]{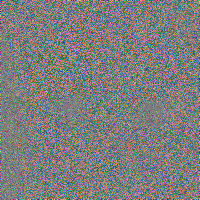}}
	{\includegraphics[width=0.6in]{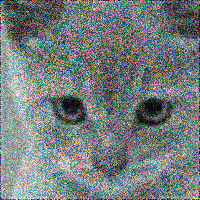}}\\
	\flushleft{\hspace{0.37in}$y$ \hspace{0.3in} ${F_\theta }{(y)^{1k}}$ \hspace{0.28in}  $\hat{n}^{1k}$ \hspace{0.32in}  $n_2^{1k}$ \hspace{0.35in}  $y_2^{1k}$  }\\
	\centering
	\vspace{0.1in}
	{\includegraphics[width=0.6in]{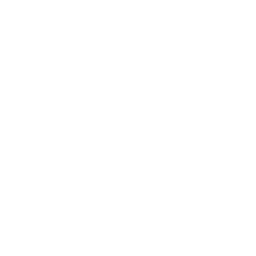}}
	{\includegraphics[width=0.6in]{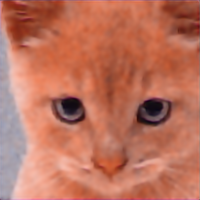}}
	{\includegraphics[width=0.6in]{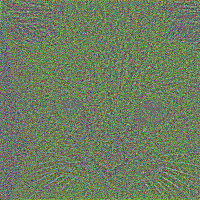}}
	{\includegraphics[width=0.6in]{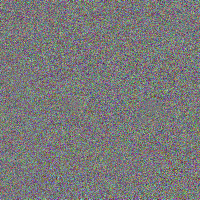}}
	{\includegraphics[width=0.6in]{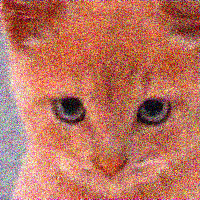}}\\
	\flushleft{\hspace{0.75in} ${F_\theta }{(y)^{20k}}$ \hspace{0.24in}  $\hat{n}^{20k}$ \hspace{0.26in}  $n_2^{20k}$ \hspace{0.32in}  $y_2^{20k}$  }\\
	\centering
	\vspace{0.1in}
	
	{\includegraphics[width=0.6in]{cat_white.png}}
	{\includegraphics[width=0.6in]{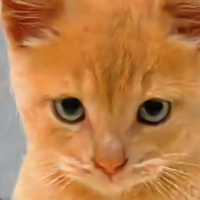}}
	{\includegraphics[width=0.6in]{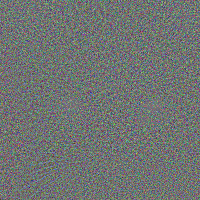}}
	{\includegraphics[width=0.6in]{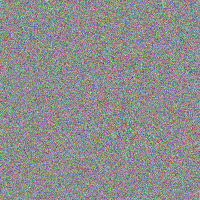}}
	{\includegraphics[width=0.6in]{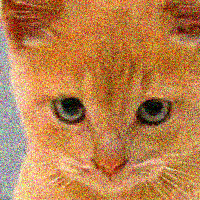}}\\
	\flushleft{\hspace{0.73in} ${F_\theta }{(y)^{300k}}$ \hspace{0.16in}  $\hat{n}^{300k}$ \hspace{0.23in}  $n_2^{300k}$ \hspace{0.28in}  $y_2^{300k}$  }\\
	\centering
	\vspace{0.1in}
	{\includegraphics[width=0.6in]{cat_white.png}}
	{\includegraphics[width=0.6in]{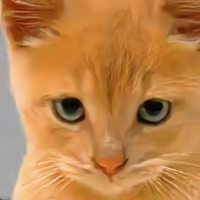}}
	{\includegraphics[width=0.6in]{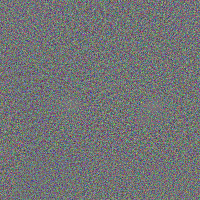}}
	{\includegraphics[width=0.6in]{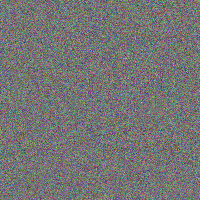}}
	{\includegraphics[width=0.6in]{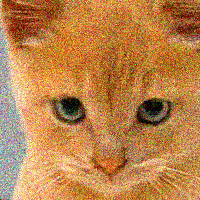}}\\
	\flushleft{\hspace{0.68in} ${F_\theta }{(y)^{1500k}}$ \hspace{0.11in}  $\hat{n}^{1500k}$ \hspace{0.17in}  $n_2^{1500k}$ \hspace{0.23in}  $y_2^{1500k}$  }\\

	\caption{Visual examples produced by SVID during training. The superscript represents the number of training steps. By multiplying with a random $m$, $\hat{n}$ is transformed into $n_2$. }
	\label{fig_cat}
\end{figure}

\subsection{Discussion of SVID}
We motivate the SVID objective of Eq. (\ref{eq.loss}) as a modification of Eq. (\ref{eq.new_reg}). From Eq. (\ref{eq.new_reg}), we observe that we can expand the self-regularization $R(y,y_2) \equiv \left\| {{F_\theta }({y_2}) - {F_\theta }(y)} \right\|_2^2$ as 
\begin{eqnarray}
\label{eq.R}
R(y,y_2) &=& \left\| {{F_\theta }({y_2}) -y + y - {F_\theta }(y)} \right\|_2^2\\
&=& \left\|{F_\theta }(y) - y \right\|_2^2 + \left\|y - {F_\theta }(y_2) \right\|_2^2\nonumber\\
&& - 2\left\langle y - {F_\theta }(y) , y-{F_\theta }(y_2)\right\rangle\nonumber .
\end{eqnarray}
We observe that the squared error $\|{F_\theta }(y) - y\|_2^2$ is implicitly included in $R(y,y_2)$, and therefore, modulo a rescaling of the regularization parameter $\alpha$, the squared error is not necessary in $Eq. (\ref{eq.new_reg})$. We also observe that the squared error term $\|y - {F_\theta }(y_2)\|_2^2$ is between $y$ and the output using $y_2$. Since the noise in $y$ and $y_2$ are assumed zero mean and independent, we can view this as an adversarial term, since it is not possible to model noise in $y$ based on the noise in $y_2$, and therefore the network defaults to their shared structure.

In Eq. (\ref{eq.R}), we need the second noisy image $y_2$. We observe that if $y_2$ is synthesized by Eq. (\ref{eq.y2}), then $y_2$ naturally has some desirable properties. Specifically,

1. Some image details may remain in $\hat{n}$. These details can be destroyed by multiplying with a random mask $m$, which makes $n_2$ a more natural-looking noise (see Figure \ref{fig_cat}).

2. $\mathbb{E}(m)=0$ leads to $\mathbb{E}(n_2)=\mathbb{E}(m)\mathbb{E}(\hat{n})=0$. Therefore, the noise of $y_2$ meets the zero-mean assumption.

\begin{figure}[t]
	\centering
	\subfigure[$1k$ iterations]{\includegraphics[width=.44\columnwidth]{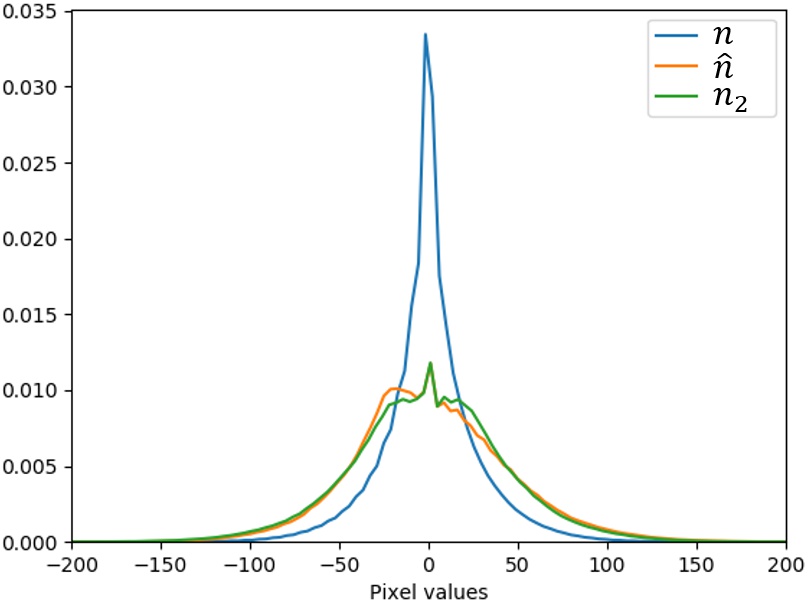}}
	\subfigure[$20k$ iterations]{\includegraphics[width=.44\columnwidth]{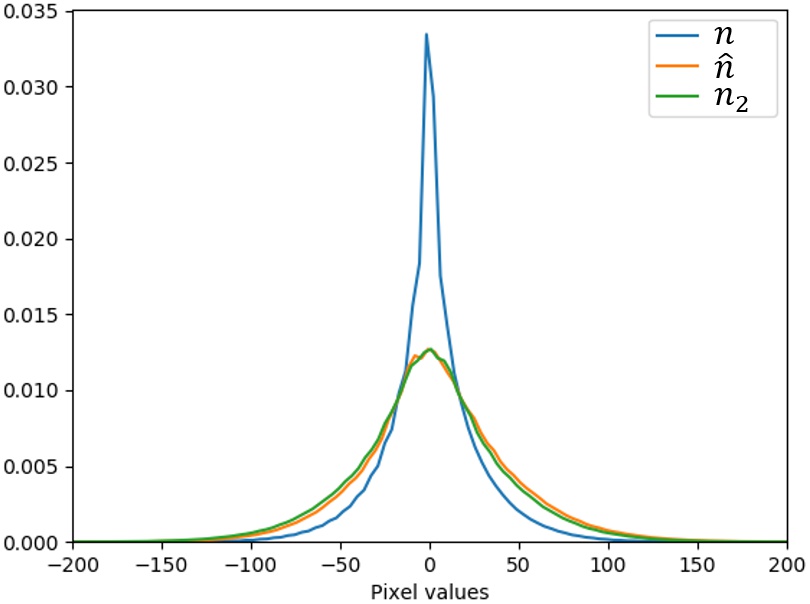}}
	\subfigure[$300k$ iterations]{\includegraphics[width=.44\columnwidth]{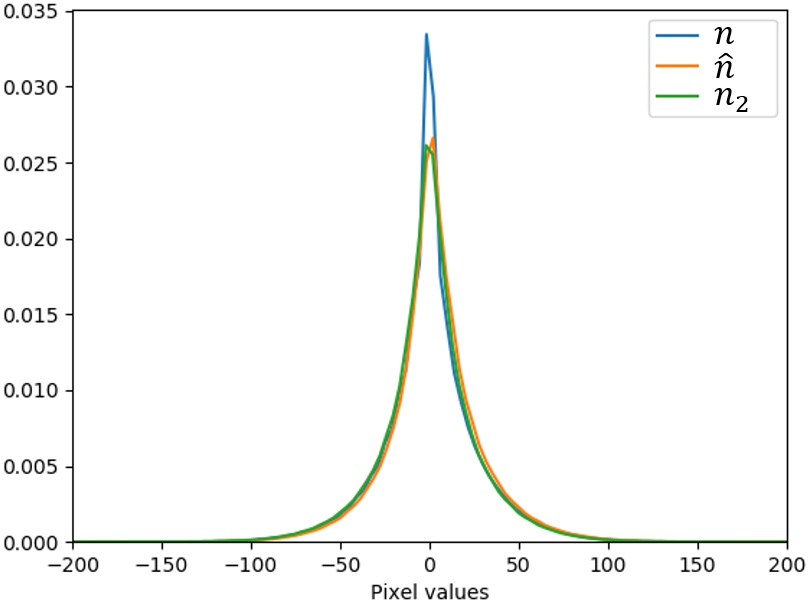}}
	\subfigure[$1500k$ iterations]{\includegraphics[width=.44\columnwidth]{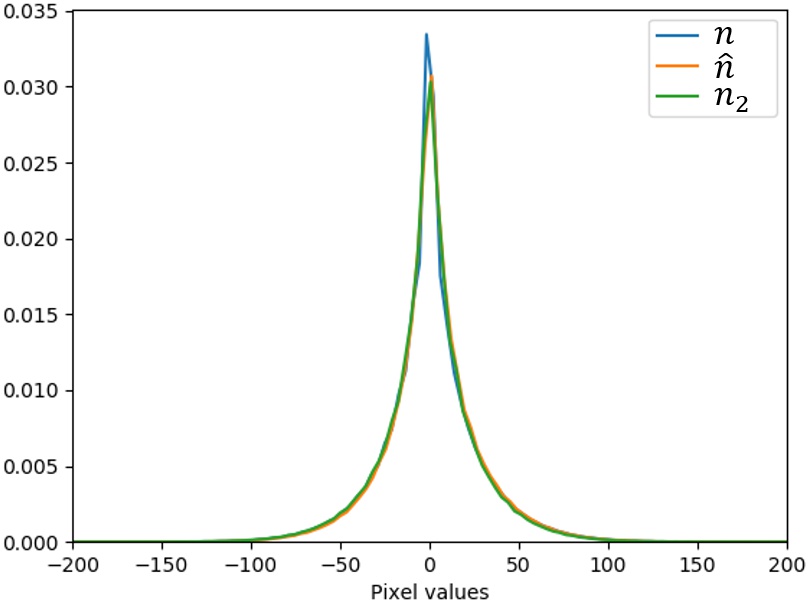}}	
	
	\caption{Statistical histograms for $20,000$ noise examples with a size of $128\times 128$. $n$ is the real noise, $\hat{n}$ and $n_2$ are the noise produced by SVID. As the training progresses, the statistical distributions of $\hat{n}$ and $n_2$ gradually converge to the distribution of $n$.}
	\label{fig_hist}
\end{figure}

3. Since $n_2$ is composed of $\hat{n}$, the statistical characteristics of $n_2$ and $\hat{n}$ should be similar, as shown in Figure \ref{fig_hist}. In addition, $m$ is random, so ${n_2} \ne \hat n$. This further leads to ${y_2} \ne y$. If ${y_2} = y$, self-verification Eq. (\ref{eq.loss}) is meaningless.

%

\begin{figure*}[t]
	\centering
	
	\subfigure[ {Clean $|$ SSIM, PSNR}]{\includegraphics[width=.23\textwidth]{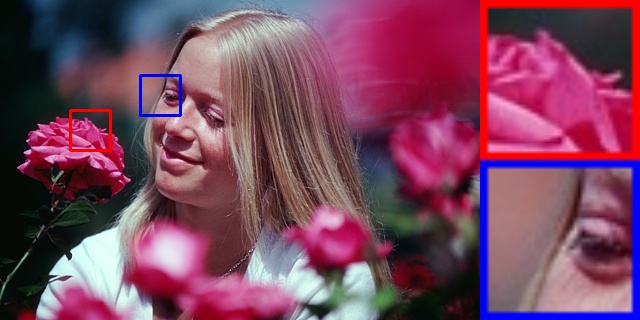}}
	\subfigure[ {Input $|$ 0.491, 24.25}]{\includegraphics[width=.23\textwidth]{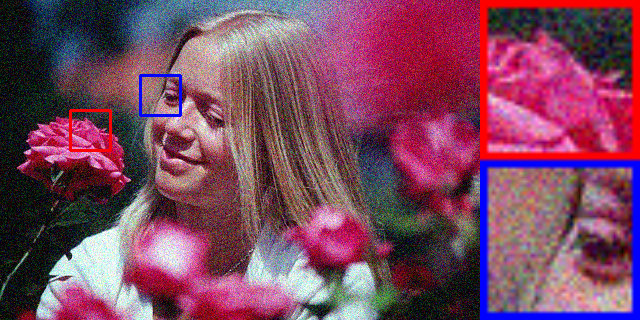}}
	\subfigure[ {BM3D $|$ 0.915, 32.70}]{\includegraphics[width=.23\textwidth]{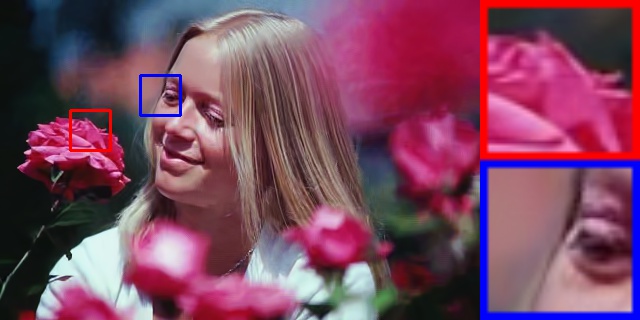}}
	\subfigure[ {DIP $|$ {0.900}, {32.00}}]{\includegraphics[width=.23\textwidth]{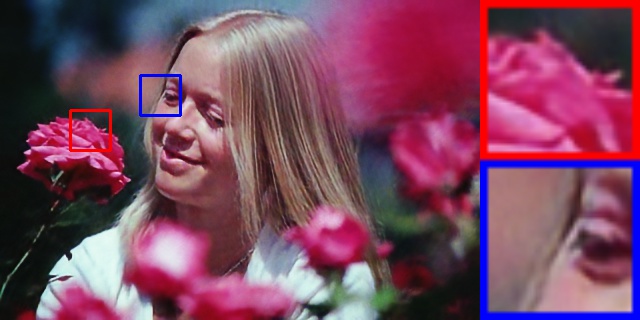}}
	\\
	\subfigure[ {N2V $|$ 0.915, 32.62}]{\includegraphics[width=.23\textwidth]{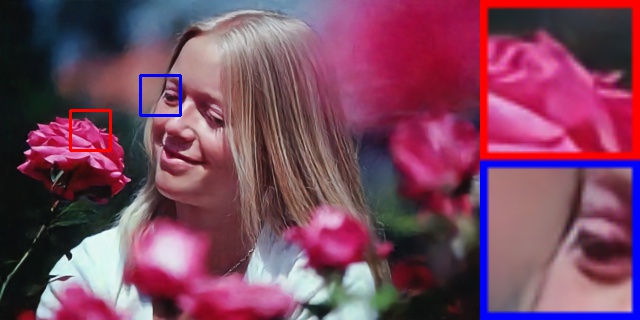}}
	\subfigure[ {Nb2Nb $|$ 0.918, 32.79}]{\includegraphics[width=.23\textwidth]{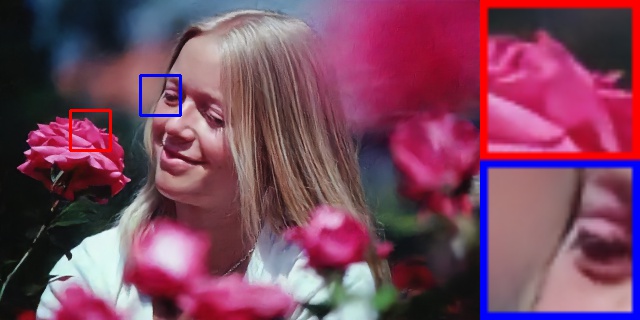}}
	\subfigure[ {U-Net $|$ \textbf{0.934}, \textbf{33.96}}]{\includegraphics[width=.23\textwidth]{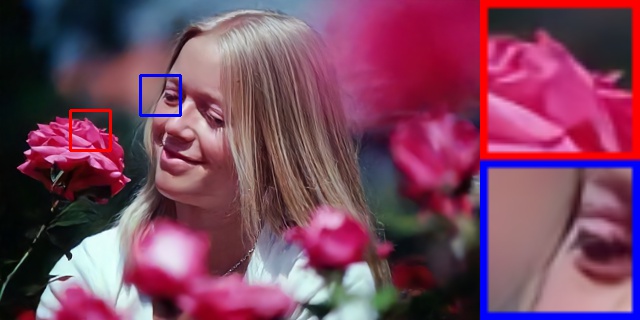}}
	\subfigure[ {Our $|$ 0.930, 33.61}]{\includegraphics[width=.23\textwidth]{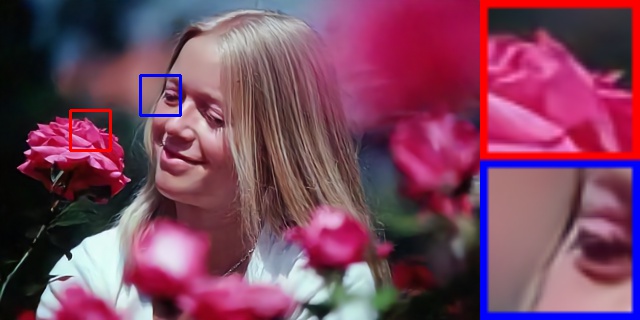}}
	
	\caption{Example results for Gaussian denoising, $\sigma=25$.}
	\label{fig_gaus}
\end{figure*} 

\begin{table*}
	\centering
	\caption{PSNR results (dB) on BSD300 data with Gaussian, Speckle and Poisson noise. \textbf{Bold}: best. \textcolor{red}{Red}: second. \textcolor{blue}{Blue}: third. We note that U-Net is a fully supervised model and can be considered an upper bound on performance since SVID also employs the U-Net framework. N2N also can be understood as a supervised model that SVID approaches. SVID outperforms all other self-supervised models and non-deep models.}
	\begin{tabular}{lrrrrrrrrrr}
		\toprule
		&Test noise level& BM3D & NLH&DIP&N2V&S2S&Nb2Nb&N2N&U-Net&SVID\\
		\midrule
		\multirow{2}*{Gaussian }&$\sigma=25$&{30.90}&30.31&{29.99}&{30.56}&29.63&{31.07}&\textcolor{red}{31.62}&\textbf{31.81}&\textcolor{blue}{31.50}\\
		
		&$\sigma \in (0, 50]$&31.69&31.77&{29.66}&{31.88}&27.76&{32.25}&\textcolor{red}{33.27}&\textbf{33.44}&\textcolor{blue}{32.97}\\
		\hline
		\multirow{2}*{Speckle }&$v=0.1$&26.64&25.18&{26.73}&{28.40}&27.60&{28.89}&\textcolor{red}{31.13}&\textbf{31.49}&\textcolor{blue}{29.73}\\
		&$v \in (0, 0.2]$&26.70&26.10&{27.06}&{28.77}&27.53&{29.29}&\textcolor{red}{31.39}&\textbf{31.86}&\textcolor{blue}{30.10}\\
		\hline
		\multirow{2}*{Poisson }&$\lambda =30$&27.70&28.49&{28.77}&29.78&29.23 &{29.97}&\textcolor{red}{30.91}&\textbf{31.09}&\textcolor{blue}{30.41}\\
		&$\lambda \in [5, 50]$&27.23&26.12&{27.98}&28.94&28.01&{29.03}&\textcolor{red}{30.25}& \textbf{30.44}&\textcolor{blue}{29.49} \\
		
		\bottomrule
	\end{tabular}
	\label{table_psnr}
\end{table*}

Our method uses a stop-gradient operation to avoid trivial solutions. We next discuss one hypothesis on why SVID is practically effective, as our experiments will show. In our method, we use the network's output ${F_\theta }(y)$ and the adaptively synthesized $y_2$ to learn denoising, hoping that the learned knowledge can adapt to the real noisy data $y$. SVID is an EM-like algorithm. It implicitly involves two sets of variables (\textit{i.e.} $\theta$ and ${F_\theta }(y)$), and solves two underlying sub-problems. The presence of the stop-gradient is the consequence of introducing the extra set of variables. 

We re-express the objective function of SVID in the following form,
\begin{equation}
\label{eq.loss2}
L(\theta ,\eta _y) = {\mathbb{E}_{y,{y_2}}}\left[ {\left\| {{F_\theta }({y_2}) - {\eta _y}} \right\|_2^2} \right].
\end{equation}

Note that Eq. (\ref{eq.loss2}) is consistent with the objective function in Eq. (\ref{eq.loss}). For clarity, we use $\eta _y$ to denote ${F_\theta }(y)$. With this formulation, we consider solving: 
\begin{equation}
\label{eq.loss3}
\mathop {\min }\limits_{\theta ,{\eta _y}} L(\theta ,{\eta _y}).
\end{equation}

The problem in Eq. (\ref{eq.loss3}) can be solved by an alternating algorithm, fixing one set of variables and solving for the other set. At each training step $t$, we randomly sample a noisy image $y$ from the given noisy dataset $\Omega^{noisy}$. Then, alternately solve the two sub-problems:
\begin{equation}
\label{eq.eta}
\eta _y^{(t)} \leftarrow {F_{{\theta ^{(t - 1)}}}}(y)
\end{equation}
\begin{equation}
\label{eq.theta}
{\theta ^{(t)}} \leftarrow \arg \mathop {\min }\limits_\theta  \left\| {{F_\theta }({y_2}) - \eta _y^{(t)}} \right\|_2^2
\end{equation}
where “$\leftarrow$” means assigning.

\begin{figure*}[t]
	\centering
	
	\subfigure[ {Clean $|$ SSIM, PSNR}]{\includegraphics[width=.23\textwidth]{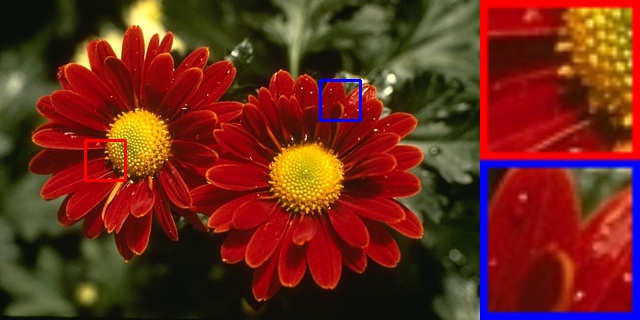}}
	\subfigure[ {Input $|$ 0.603, 23.99}]{\includegraphics[width=.23\textwidth]{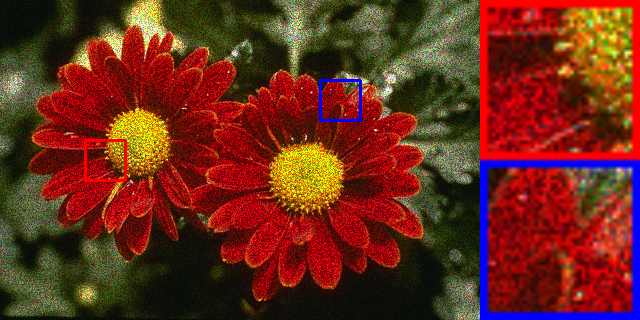}}
	\subfigure[ {BM3D $|$ 0.860, 28.96}]{\includegraphics[width=.23\textwidth]{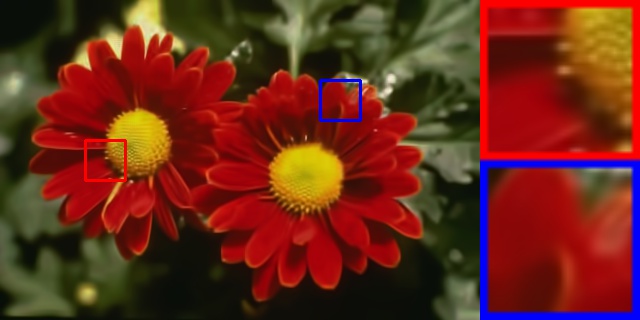}}
	\subfigure[ {DIP $|$ {0.868}, {29.67}}]{\includegraphics[width=.23\textwidth]{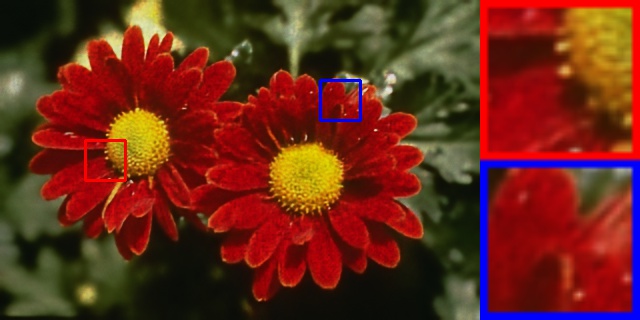}}
	\\
	\subfigure[ {N2V $|$ 0.937, 32.28}]{\includegraphics[width=.23\textwidth]{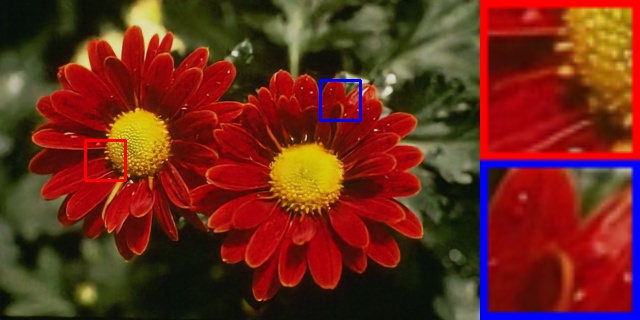}}
	\subfigure[ {Nb2Nb $|$ 0.952, 32.64}]{\includegraphics[width=.23\textwidth]{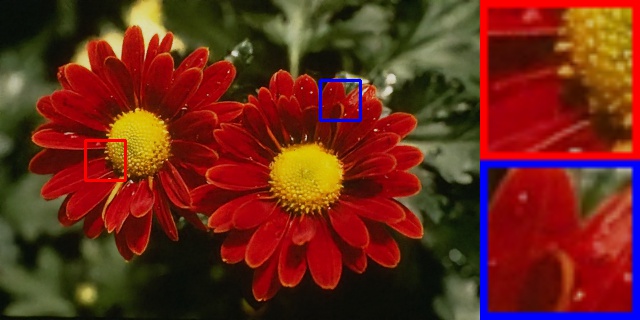}}
	\subfigure[ {U-Net $|$ \textbf{0.966}, \textbf{35.55}}]{\includegraphics[width=.23\textwidth]{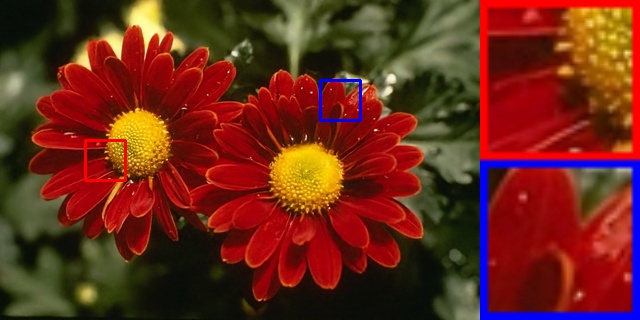}}
	\subfigure[ {Our $|$ 0.956, 33.80}]{\includegraphics[width=.23\textwidth]{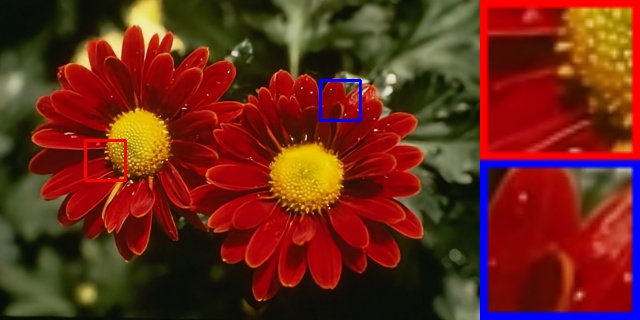}}
	
	\caption{Example results for Speckle denoising, $v=0.1$.}
	\label{fig_spk}
\end{figure*} 

\begin{figure*}[t]
	\centering
	
	\subfigure[ {Clean $|$ SSIM, PSNR}]{\includegraphics[width=.23\textwidth]{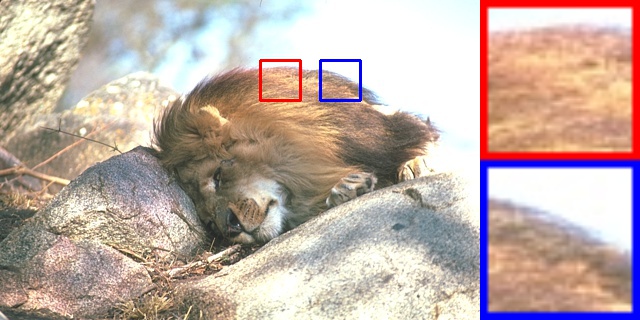}}
	\subfigure[ {Input $|$ 0.486, 20.71}]{\includegraphics[width=.23\textwidth]{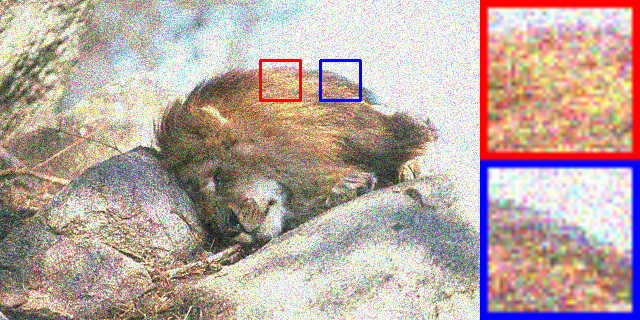}}
	\subfigure[ {BM3D $|$ 0.723, 24.15}]{\includegraphics[width=.23\textwidth]{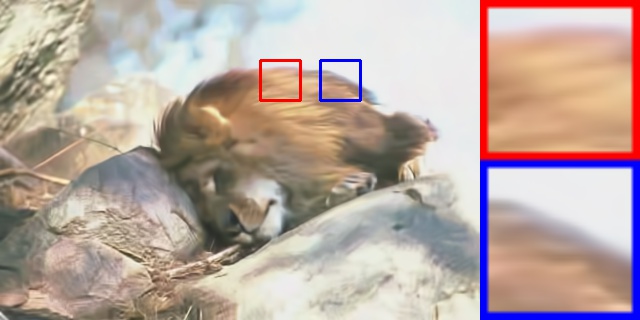}}
	\subfigure[ {DIP $|$ {0.790}, {24.80}}]{\includegraphics[width=.23\textwidth]{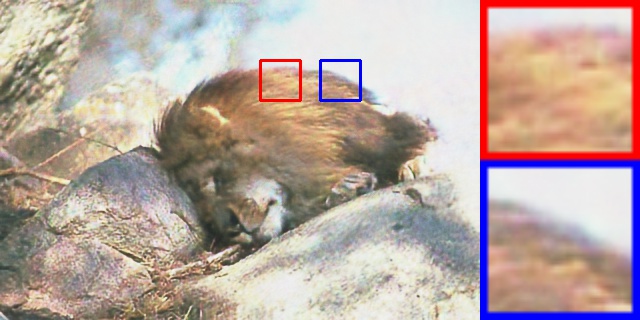}}
	\\
	\subfigure[ {N2V $|$ 0.821, 25.53}]{\includegraphics[width=.23\textwidth]{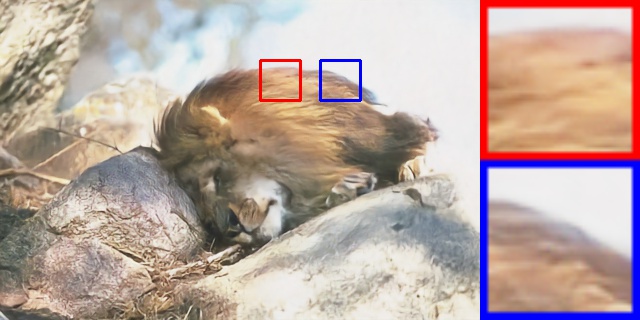}}
	\subfigure[ {Nb2Nb $|$ 0.832, 25.47}]{\includegraphics[width=.23\textwidth]{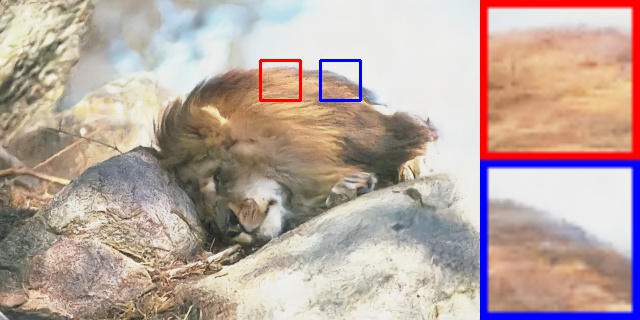}}
	\subfigure[ {U-Net $|$ \textbf{0.852}, \textbf{28.15}}]{\includegraphics[width=.23\textwidth]{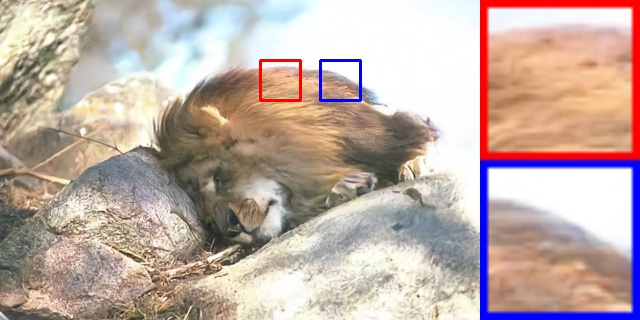}}
	\subfigure[ {Our $|$ 0.847, 27.93}]{\includegraphics[width=.23\textwidth]{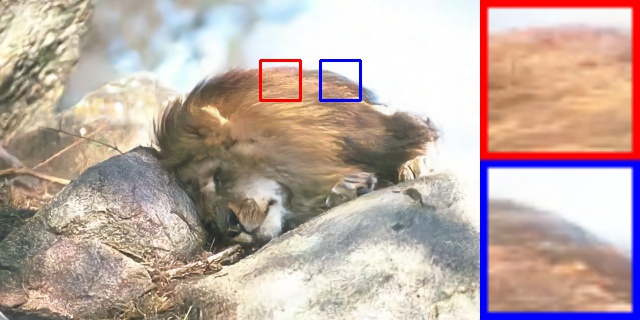}}
	
	\caption{Example results for Poisson denoising, $\lambda = 30$.}
	\label{fig_pos}
\end{figure*}

\textbf{\textit{Solving for $\eta _y$.}} There is no optimization in this part, but simply a redefinition of the term $\eta _y$ to be the output of the neural network on $y$. Moreover, previous work on DIP has revealed that the CNN naturally tends to capture low-level image statistics. Therefore, the similarity between $\eta _y^{(t)}$ and $x$ will gradually increase.

\textbf{\textit{Solving for $\theta$.}} $y_2$ is constructed by Eq. (\ref{eq.y2}) using $\eta _y^{(t)}$. The more similar $\eta _y^{(t)}$ is to $x$, the closer the noise statistics of $y_2$ and $y$ are (see Figure \ref{fig_hist} for illustration). The denoising network is improved by solving Eq. (\ref{eq.theta}). $\eta _y^{(t)}$ is fixed in this subproblem, so the gradient does not backpropagate to $\eta _y^{(t)}$. 

In short, one output of the network is obtained at training step $t$, which serves as a ``prior'' target for denoising. $\eta _y^{(t)}$ contains the low-level image statistics required to restore a clean image. Therefore, we use $\eta _y^{(t)}$ as a label to improve the denoising ability of the network. This leads to the network to produce a better prior at time $t+1$. By solving the two sub-problems alternately at each training step, ${F_\theta }(y)$ approaches the true underlying $x$.

\subsection{Training Details}

The denoising CNN in SVID is a simple U-Net \cite{ronneberger2015u}. We use PyTorch and Adam \cite{kingma2014adam} with a batch size of 1 to train the network. The training images are randomly cropped into $128 \times 128$ patches before being input to the network. The learning rate is fixed to $0.0002$  for the first $1,000,000$ iterations and linearly decays to 0 for the next $1,000,000$ iterations.\footnote{We will include a reference here to our code and datasets.} After training, the CNN with fixed parameters $\theta^*$ is directly applied to other noisy images.


\section{Experiments}
We evaluate SVID on various denoising tasks.

\subsection{Synthetic Noise}
We collected 4744 images from the Waterloo Exploration Database \cite{ma2016waterloo} to synthesize noisy images for training. Several state-of-the-art denoising methods are adopted for performance comparison, including model-based methods BM3D \cite{dabov2007image} and NLH \cite{hou2020nlh}, self-learning methods DIP \cite{ulyanov2018deep}, Noise2Void(N2V) \cite{krull2019noise2void}, Self2Self(S2S) \cite{quan2020self2self} and Neighbor2Neighbor(Nb2Nb) \cite{huang2021neighbor2neighbor}, other deep learning methods include Noise2Noise(N2N) \cite{lehtinen2018noise2noise} and a common fully-supervised U-Net. U-Net is trained with noisy/clean image pairs, while N2N uses paired noisy/noisy images. For the sake of fairness, N2N, U-Net and our SVID adopt the same network architecture to perform denoising. BSD300 \cite{martin2001database} is the test set of the following experiments.

\begin{figure}[t]
	\centering
	\subfigure[Clean (Normal-dose) $|$ \color{white}{a} \color{black} SSIM, PSNR]{\includegraphics[width=.48\columnwidth]{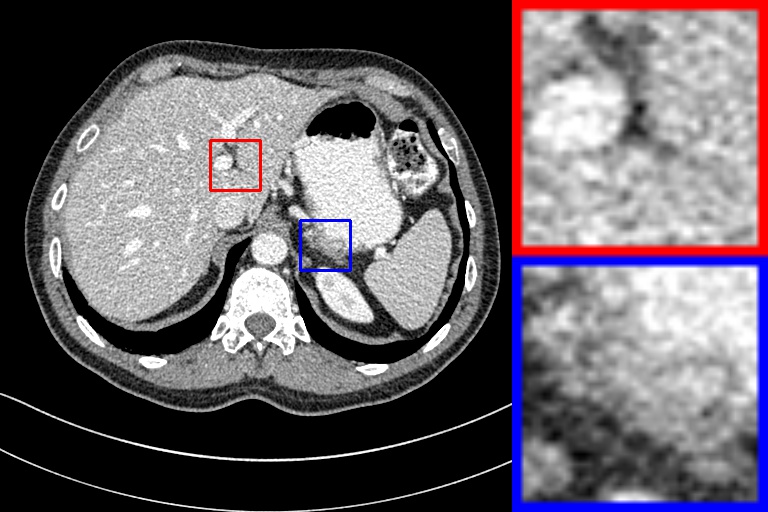}}
	\subfigure[Noisy (Low-dose)$|$ \color{white}{aa} \color{black} 0.724, 19.55  ]{\includegraphics[width=.48\columnwidth]{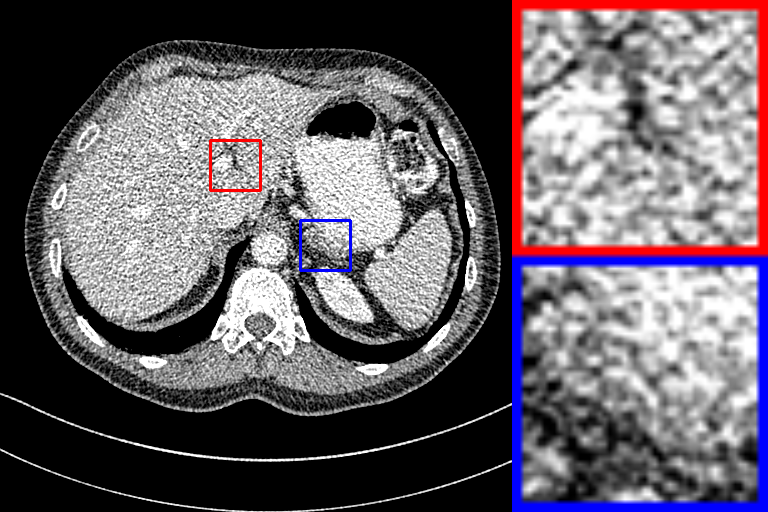}}\\
	\subfigure[U-Net $|$ 0.788, 25.80]{\includegraphics[width=.48\columnwidth]{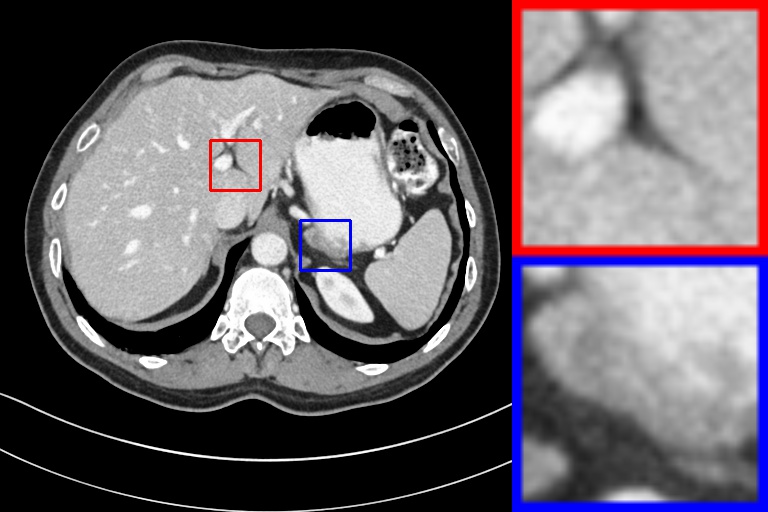}}
	\subfigure[SVID $|$ 0.779, 25.01]{\includegraphics[width=.48\columnwidth]{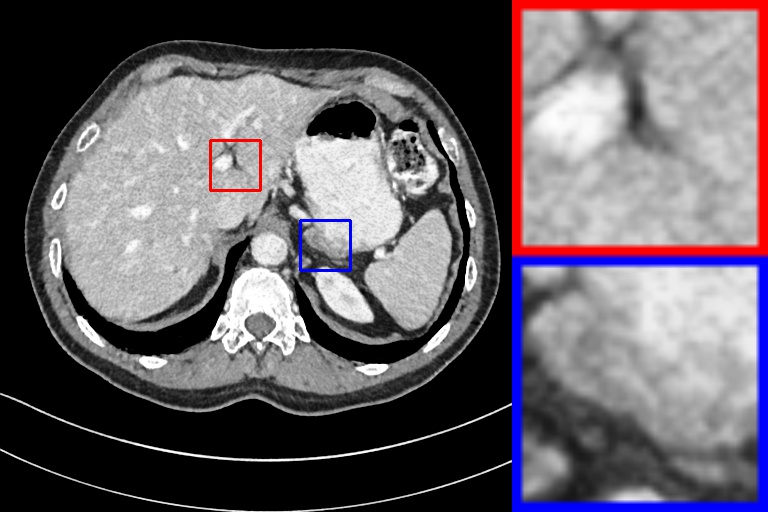}}

	\caption{Low-dose CT denoising example. }
	\label{fig_ct}
\end{figure}

%
%

%
%
%
%
%
%

\textbf{\textit{Gaussian noise.}} We first study the effectiveness of SVID on additive Gaussian noise. Each training example is corrupted by zero-mean Gaussian noise with a random standard deviation $\sigma \in \left( 0,50\right]$. For testing, we synthesize two sets of noisy images, using a fixed noise level $\sigma = 25$ and a variable $\sigma \in \left( 0,50\right]$. We present the quantitative results in Table \ref{table_psnr}. Visual comparisons can be found in Figure \ref{fig_gaus}. In addition, Figure \ref{fig_hist} shows the statistical histograms of the noise removed by SVID during training.

\textbf{\textit{Speckle noise.}} Multiplicative speckle noise is signal dependent and is often observed in medical ultrasonic images and radar images. The speckle noise in the image is modeled as the random value multiplied by the pixel value of the latent signal $x$, which can be expressed as $y=x+x\cdot n$. In this model, $n$ is an uniform noise with a mean of $0$ and a variance of $v$. We vary the noise variance $v \in \left(0,0.2 \right]$ during training. The results and the comparisons of denoising on Speckle noise can be seen in Table \ref{table_psnr} and Figure \ref{fig_spk}.

\textbf{\textit{Poisson noise.}} In our third experiment we consider Poisson noise, which can be used to model photon noise in imaging sensors. Poisson noise is also signal-dependent because its expected magnitude depends on the pixel brightness. We randomize the noise magnitude $\lambda \in \left[5,50\right]$ separately for each training example. Quantitative and subjective comparisons are reported in Table \ref{table_psnr} and Figure \ref{fig_pos}, respectively.

\textbf{\textit{Discussion.}} In the above comparisons, SVID shows excellent denoising results both subjectively and quantitatively. These results demonstrate that the output of the denoising network can be a prior for denoising, as this paper conjectures. The denoising performance of SVID outperforms other self-supervised methods (\textit{i.e.} DIP, N2V, S2S, Nb2Nb) by a large margin, which confirms the effectiveness of self-verification. In SVID, the denoising ability of the network can be improved by self-verification. In addition, SVID is better than model-based methods BM3D and NLH. This is because the deep image prior captured by the network can better model the properties of natural images than the traditional non-local self-similarity prior. We also noticed that SVID is inferior to supervised methods U-Net and N2N, which is reasonable considering that our SVID has no access to paired data for supervision. In short, our SVID consistently exhibits encouraging denoising performance for various types of noise. The denoised images are clean and sharp. More importantly, SVID does not rely on paired data or pre-known noise statistics, which shows its potential value for many practical applications.

\begin{table}
	\centering
	\caption{Quantitative results for low-dose CT denoising. SVID approaches the fully supervised U-Net.}
	\begin{tabular}{lrrr}
		\toprule
		{} & Noisy &U-Net&SVID\\
		\midrule
		SSIM&0.766&0.827&0.813\\
		PSNR&22.61&28.56&27.20\\

		\bottomrule
	\end{tabular}
	\label{table_CT}
\end{table}
\subsection{Low-Dose CT Denoising}
Computed tomography (CT) is a popular imaging modality in clinical diagnosis. Despite the healthcare benefits, the extensive use of CT has caused public concern about the potential risks of X-ray radiation. Reducing radiation dose is an effective way to deal with this problem. However, a decrease of radiation dose is associated with an increase of noise and artifacts in the reconstructed image, which may adversely affect subsequent diagnosis. Therefore, denoising is a critical post-processing step for low-dose CT images \cite{shan2019competitive}. Since the noise statistics in CT images are very complicated and paired training data is difficult to obtain, the self-supervised denoising methods can better cater to low-dose CT images. We further evaluate SVID in the task of low-dose CT denoising.

We use a real clinical dataset authorized for the 2016 NIH-AAPM-Mayo Clinic LDCT Grand Challenge by Mayo Clinic.\footnote{https://www.aapm.org/GrandChallenge/LowDoseCT/} This dataset contains 5946 pairs of normal-/low- dose images with a slice thickness of 1mm. We randomly select 5000 pairs of images as the training set, and the rest as the test set. We compare SVID with a fully supervised U-Net. Note that SVID only uses low-dose CT images to learn denoising without the normal-dose images. The comparison results are reported in Figure \ref{fig_ct} and Table \ref{table_CT}. As can be seen, our SVID is also effective for noise in CT. SVID is successful in removing the majority of noise and restoring high-quality image details. Its performance is close to the fully supervised U-Net. 



\begin{figure}[t]
	\centering
	\subfigure[{Clean}]{\includegraphics[width=.48\columnwidth]{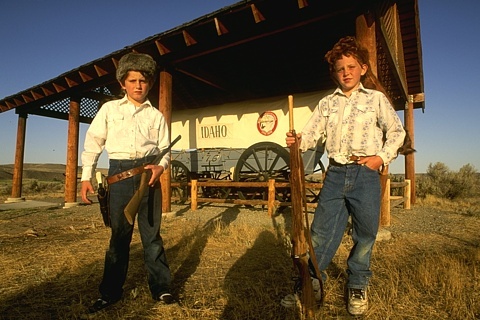}}
	\subfigure[{Input}]{\includegraphics[width=.48\columnwidth]{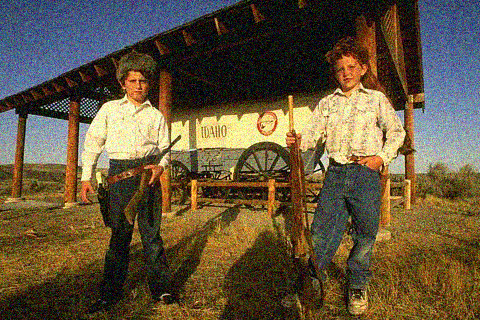}}\\
	\subfigure[{w/o stop-gradient}]{\includegraphics[width=.48\columnwidth]{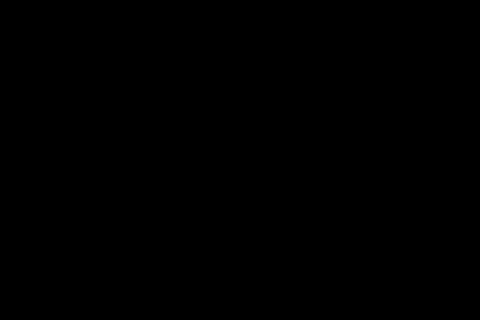}}
	\subfigure[{w/ stop-gradient}]{\includegraphics[width=.48\columnwidth]{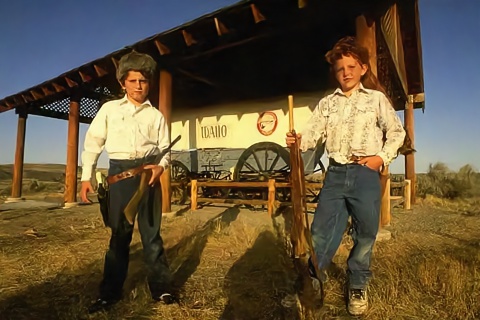}}

	\caption{SVID with and without stop-gradient. The noise in (b) is Gaussian ($\sigma=25$).}
	\label{fig_ablation}
\end{figure}

\subsection{Stop-gradient}
Our SVID employs a Siamese network to perform self-verification. An undesired trivial solution to Siamese networks is all outputs collapse to a constant. There are several strategies to prevent Siamese networks from collapsing, such as contrastive learning, clustering and stop-gradients \cite{chen2021exploring}. In SVID, the stop-gradient operation is a natural consequence of introducing an extra variable. Here, our aim is to show that the stop-gradient is critical to preventing SVID from collapsing. 

Figure \ref{fig_ablation} presents a comparison with and without using the stop-gradient. Without stop-gradients, all outputs of the denoising network collapse to 0 and the objective function Eq. (\ref{eq.loss2}) reaches its minimum of 0. With stop-gradients, the objective function Eq. (\ref{eq.loss2}) implicitly constrains the removed noise to be zero-mean. Therefore, the denoised image is prevented from collapsing to 0. Moreover, the deep CNN tends to capture low-level image statistics, which promotes the denoised image to look like a clean natural image.

\section{Conclusion}
We have demonstrated that the output of a neural network can be a prior for image denoising. Based on this prior, we developed a new self-verification regularization where even without any clean data, the network can learn to denoise in a self-supervised manner. We demonstrated the effectiveness and broad applicability of the proposed method on several denoising tasks involving different noise properties and data types. We believe that this study hints at the advantage of exploring the use of CNNs to generate deep image priors for various CV tasks.

{\small
	\bibliographystyle{ieee}
	\bibliography{egbib}

\begin{thebibliography}{10}\itemsep=-1pt

\bibitem{batson2019noise2self}
J.~Batson and L.~Royer.
\newblock Noise2self: Blind denoising by self-supervision.
\newblock In {\em ICML}, 2019.

\bibitem{buades2005non}
A.~Buades, B.~Coll, and J.-M. Morel.
\newblock A non-local algorithm for image denoising.
\newblock In {\em CVPR}, 2005.

\bibitem{chen2021exploring}
X.~Chen and K.~He.
\newblock Exploring simple siamese representation learning.
\newblock In {\em CVPR}, 2021.

\bibitem{dabov2007image}
K.~Dabov, A.~Foi, V.~Katkovnik, and K.~Egiazarian.
\newblock Image denoising by sparse 3-d transform-domain collaborative
  filtering.
\newblock {\em IEEE Transactions on Image Processing}, 16(8):2080--2095, 2007.

\bibitem{dong2015low}
W.~Dong, G.~Li, G.~Shi, X.~Li, and Y.~Ma.
\newblock Low-rank tensor approximation with laplacian scale mixture modeling
  for multiframe image denoising.
\newblock In {\em ICCV}, 2015.

\bibitem{grill2020bootstrap}
J.-B. Grill, F.~Strub, F.~Altch{\'e}, C.~Tallec, P.~H. Richemond,
  E.~Buchatskaya, C.~Doersch, B.~A. Pires, Z.~D. Guo, M.~G. Azar, et~al.
\newblock Bootstrap your own latent: A new approach to self-supervised
  learning.
\newblock In {\em NeurIPS}, 2020.

\bibitem{gu2014weighted}
S.~Gu, L.~Zhang, W.~Zuo, and X.~Feng.
\newblock Weighted nuclear norm minimization with application to image
  denoising.
\newblock In {\em CVPR}, 2014.

\bibitem{hou2020nlh}
Y.~Hou, J.~Xu, M.~Liu, G.~Liu, L.~Liu, F.~Zhu, and L.~Shao.
\newblock Nlh: A blind pixel-level non-local method for real-world image
  denoising.
\newblock {\em IEEE Transactions on Image Processing}, 29:5121--5135, 2020.

\bibitem{huang2021neighbor2neighbor}
T.~Huang, S.~Li, X.~Jia, H.~Lu, and J.~Liu.
\newblock Neighbor2neighbor: Self-supervised denoising from single noisy
  images.
\newblock In {\em CVPR}, 2021.

\bibitem{kingma2014adam}
D.~P. Kingma and J.~Ba.
\newblock Adam: A method for stochastic optimization.
\newblock {\em arXiv preprint arXiv:1412.6980}, 2014.

\bibitem{krull2019noise2void}
A.~Krull, T.-O. Buchholz, and F.~Jug.
\newblock Noise2void-learning denoising from single noisy images.
\newblock In {\em CVPR}, 2019.

\bibitem{laine2019high}
S.~Laine, T.~Karras, J.~Lehtinen, and T.~Aila.
\newblock High-quality self-supervised deep image denoising.
\newblock In {\em NeurIPS}, 2019.

\bibitem{lehtinen2018noise2noise}
J.~Lehtinen, J.~Munkberg, J.~Hasselgren, S.~Laine, T.~Karras, M.~Aittala, and
  T.~Aila.
\newblock Noise2noise: Learning image restoration without clean data.
\newblock In {\em ICML}, 2018.

\bibitem{liu2018non}
D.~Liu, B.~Wen, Y.~Fan, C.~C. Loy, and T.~S. Huang.
\newblock Non-local recurrent network for image restoration.
\newblock In {\em NeurIPS}, 2018.

\bibitem{ma2016waterloo}
K.~Ma, Z.~Duanmu, Q.~Wu, Z.~Wang, H.~Yong, H.~Li, and L.~Zhang.
\newblock Waterloo exploration database: New challenges for image quality
  assessment models.
\newblock {\em IEEE Transactions on Image Processing}, 26(2):1004--1016, 2016.

\bibitem{martin2001database}
D.~Martin, C.~Fowlkes, D.~Tal, and J.~Malik.
\newblock A database of human segmented natural images and its application to
  evaluating segmentation algorithms and measuring ecological statistics.
\newblock In {\em ICCV}, 2001.

\bibitem{meng2013robust}
D.~Meng and F.~De~La~Torre.
\newblock Robust matrix factorization with unknown noise.
\newblock In {\em ICCV}, 2013.

\bibitem{quan2020self2self}
Y.~Quan, M.~Chen, T.~Pang, and H.~Ji.
\newblock Self2self with dropout: Learning self-supervised denoising from
  single image.
\newblock In {\em CVPR}, 2020.

\bibitem{ronneberger2015u}
O.~Ronneberger, P.~Fischer, and T.~Brox.
\newblock U-net: Convolutional networks for biomedical image segmentation.
\newblock In {\em MICCAI}, 2015.

\bibitem{selesnick2017total}
I.~Selesnick.
\newblock Total variation denoising via the moreau envelope.
\newblock {\em IEEE Signal Processing Letters}, 24(2):216--220, 2017.

\bibitem{shan2019competitive}
H.~Shan, A.~Padole, F.~Homayounieh, U.~Kruger, R.~D. Khera, C.~Nitiwarangkul,
  M.~K. Kalra, and G.~Wang.
\newblock Competitive performance of a modularized deep neural network compared
  to commercial algorithms for low-dose ct image reconstruction.
\newblock {\em Nature Machine Intelligence}, 1(6):269--276, 2019.

\bibitem{simoncelli1996noise}
E.~P. Simoncelli and E.~H. Adelson.
\newblock Noise removal via bayesian wavelet coring.
\newblock In {\em ICIP}, 1996.

\bibitem{ulyanov2018deep}
D.~Ulyanov, A.~Vedaldi, and V.~Lempitsky.
\newblock Deep image prior.
\newblock In {\em CVPR}, 2018.

\bibitem{xu2018trilateral}
J.~Xu, L.~Zhang, and D.~Zhang.
\newblock A trilateral weighted sparse coding scheme for real-world image
  denoising.
\newblock In {\em ECCV}, 2018.

\bibitem{yao2019nonconvex}
J.~Yao, D.~Meng, Q.~Zhao, W.~Cao, and Z.~Xu.
\newblock Nonconvex-sparsity and nonlocal-smoothness-based blind hyperspectral
  unmixing.
\newblock {\em IEEE Transactions on Image Processing}, 28(6):2991--3006, 2019.

\bibitem{yue2019variational}
Z.~Yue, H.~Yong, Q.~Zhao, D.~Meng, and L.~Zhang.
\newblock Variational denoising network: Toward blind noise modeling and
  removal.
\newblock In {\em NeurIPS}, 2019.

\bibitem{yue2020dual}
Z.~Yue, Q.~Zhao, L.~Zhang, and D.~Meng.
\newblock Dual adversarial network: Toward real-world noise removal and noise
  generation.
\newblock In {\em ECCV}, 2020.

\bibitem{zbontar2021barlow}
J.~Zbontar, L.~Jing, I.~Misra, Y.~LeCun, and S.~Deny.
\newblock Barlow twins: Self-supervised learning via redundancy reduction.
\newblock {\em arXiv preprint arXiv:2103.03230}, 2021.

\bibitem{zhang2017beyond}
K.~Zhang, W.~Zuo, Y.~Chen, D.~Meng, and L.~Zhang.
\newblock Beyond a gaussian denoiser: Residual learning of deep cnn for image
  denoising.
\newblock {\em IEEE Transactions on Image Processing}, 26(7):3142--3155, 2017.

\bibitem{zhang2018ffdnet}
K.~Zhang, W.~Zuo, and L.~Zhang.
\newblock Ffdnet: Toward a fast and flexible solution for cnn-based image
  denoising.
\newblock {\em IEEE Transactions on Image Processing}, 27(9):4608--4622, 2018.

\bibitem{zhang2021accurate}
Y.~Zhang, K.~Li, K.~Li, G.~Sun, Y.~Kong, and Y.~Fu.
\newblock Accurate and fast image denoising via attention guided scaling.
\newblock {\em IEEE Transactions on Image Processing}, 30:6255--6265, 2021.

\bibitem{zhang2019residual}
Y.~Zhang, K.~Li, K.~Li, B.~Zhong, and Y.~Fu.
\newblock Residual non-local attention networks for image restoration.
\newblock In {\em ICLR}, 2019.

\bibitem{zhang2020residual}
Y.~Zhang, Y.~Tian, Y.~Kong, B.~Zhong, and Y.~Fu.
\newblock Residual dense network for image restoration.
\newblock {\em IEEE Transactions on Pattern Analysis and Machine Intelligence},
  43(7):2480--2495, 2020.

\bibitem{zhao2014robust}
Q.~Zhao, D.~Meng, Z.~Xu, W.~Zuo, and L.~Zhang.
\newblock Robust principal component analysis with complex noise.
\newblock In {\em ICML}, 2014.

\end{thebibliography}
}

\end{document}